\documentclass[english,useAMS, usenatbib]{mn2e}
\usepackage[T1]{fontenc}
\usepackage[latin9]{inputenc}
\usepackage{units}
\usepackage{rotating}
\usepackage{url}
\usepackage{amsmath}
\usepackage{commath}
\usepackage{amssymb}
\usepackage{graphicx}
\usepackage[caption=false]{subfig}
\usepackage{pdfpages}
\usepackage{esint}
\usepackage[authoryear]{natbib}
\usepackage{listings}
\usepackage{textcomp}
\usepackage[export]{adjustbox}

\title[Formation and coalescence sites of the first GW events]{The formation and coalescence sites of the first gravitational wave events}
\author[Schneider et al.]{Raffaella Schneider$^{1,2,3}$\thanks{E-mail:
raffaella.schneider@oa-roma.inaf.it}, Luca Graziani$^{2}$, Stefania Marassi$^{1, 2}$, Mario Spera$^{4,5}$\newauthor 
and Michela Mapelli$^{4,5}$, Matteo Alparone$^{1,2}$, Matteo de Bennassuti$^{1,2}$ \\
$^{1}$Dipartimento di Fisica, ``Sapienza'' Universit$\grave{a}$ di Roma, Piazzale Aldo Moro 5, 00185, Roma, Italy \\
$^{2}$INAF/Osservatorio Astronomico di Roma, Via di Frascati 33, 00040 Monte Porzio Catone, Italy\\
$^{3}$INFN, Sezione Roma1, Dipartimento di Fisica, ``Sapienza'' Universit$\grave{a}$ di Roma, Piazzale Aldo Moro 5, 00185, Roma, Italy \\
$^{4}$INAF/Osservatorio Astronomico di Padova, Vicolo dell' Osservatorio 5, I-35122, Padova, Italy \\
$^{5}$INFN, Sezione di Milano-Bicocca, Piazza della Scienza 3, 20126 Milano, Italy}

\begin{document}

\date{24 July 2017}

\pagerange{\pageref{firstpage}--\pageref{lastpage}} \pubyear{2017}

\maketitle
\label{firstpage}

\begin{abstract}
We present a novel theoretical model to characterize the formation and coalescence sites of compact binaries in a cosmological context.
This is based on the coupling between the binary population synthesis code \texttt{SeBa}
with a simulation following the formation of a Milky Way-like halo in a well resolved cosmic
volume of 4 cMpc, performed with the \texttt{GAMESH} pipeline.
We have applied this technique to investigate when and where systems with properties similar to the recently observed LIGO/VIRGO events
are more likely to form and where they are more likely to reside when they coalesce.
We find that more than 70\% of GW151226 and LVT151012-like systems form in galaxies with stellar mass $M_{\ast} > 10^8 M_\odot$
in the redshift range [0.06 - 3] and [0.14 - 11.3], respectively. All GW150914-like systems form
in low-metallicity dwarfs with $M_\ast < 5 \times 10^6 M_\odot$ at $2.4 \leq z \leq 4.2$.
Despite these initial differences, by the time they reach coalescence
the observed events are most likely hosted by star forming galaxies with $M_\ast > 10^{10} M_\odot$.
Due to tidal stripping and radiative feedback, a non negligible fraction of GW150914-like candidates end-up in
galaxies with properties similar to dwarf spheroidals and ultra-faint satellites.
\end{abstract}


\section{Introduction}

The recent detection of gravitational waves from the LIGO/VIRGO collaboration has opened the new 
era of gravitational wave astronomy \citep{ABBOTT2016_0914, ABBOTT2016_1226}. 
During the first Advanced LIGO observing run (hereafter O1, from September 12 2015, to January
19 2016), two sources have been unambiguously detected (GW150914 and GW151226), while a third one 
(LVT151012) was below detection threshold, but with 87\% probability of being of astrophysical origin
\citep{ABBOTT2016_BHBHO1}. At the time of submission of this letter, the detection of a fourth event has been
reported, GW170104\footnote{The analysis presented is limited to the first three events.}\citep{ABBOTT2017}.
 
All the four sources are powered by the inspiral and coalescence of two 
black holes (BHs) and the binary properties, estimated from the gravitational wave signals, have profound astrophysical implications 
\citep{ABBOTT2016_Astro}.
In particular, GW150914 originated from the merger of two $\sim 30 \, M_\odot$ BHs at $z \sim 0.09$, 
suggesting that stellar BHs with such large masses can form in nature, evolve in a binary system, and coalesce within a Hubble time. Comparison with stellar and
binary evolution models suggests progenitor stars with low metallicity, $Z < 0.5 Z_\odot$  \citep{ABBOTT2016_0914, BELCZYNSKI2016, MAPELLI2016}.
In fact, reduced mass loss at low metallicity favors the formation of more massive stellar remnants \citep{MAPELLI2009, MAPELLI2010, BELCZYNSKI2010, SPERA2015}. 

The limit on the metallicity potentially provides very interesting constraints on the birth environment of massive BH binaries.
If the binary BH 
system has merged in a short time, it must have formed in a rare, low-metallicity
dwarf galaxy. Alternatively, in the long merger time scenario ($> 9$ Gyr),
it could have formed at high redshift ($z > 2$), where low-metallicity star formation is
expected to be more common  \citep{ ABBOTT2016_Astro}.

The main goal of this paper is to investigate when and where massive BH binary
systems with properties similar to GW150914, GW151226, and LVT151012 are more likely to form, and 
where they are more likely to reside at the time of their coalescence.

Following its discovery, several attempts have been made to determine the environment in which GW150914
formed. \citet{BELCZYNSKI2016} and \citet{Dvorkin2016} use different massive BH formation scenarios
and the observationally inferred cosmic star 
formation rate density evolution with a metallicity-dependent correction to estimate the redshift evolution 
of birth and merger rate of massive BH binaries. Although these studies may provide useful constraints 
on the redshift dependent birth and merger histories of compact
binary systems, \citep{SCHNEIDER2001, REGIMBAU2011, MARASSI2011,DOMINIK2013}, 
they are not capable of discriminating individual formation or coalescence sites. 
Using observationally inferred galaxy scaling relations \citet{LAMBERTS2016} and \citet{Elbert2017}
predict the low redshift BH binary merger rate as a function of the present-day host galaxy mass. 
They find that GW150914-like events with short merger timescales would
be primarily localized in dwarf galaxies, and in massive galaxies otherwise. 
Conversely, using cosmological simulations of galaxies with
different masses at $z=0$, \citet{OSHAUGHNESSY2016}  find that a 
present-day dwarf galaxy can host a considerably larger BH binary merger rate compared to a massive galaxy.
However, some of the galaxy scaling relations
adopted by \citet{LAMBERTS2016} are not directly observed for low-mass faint galaxies,
and are therefore based on extrapolations. This may introduce considerable uncertainties in the estimated properties
of the most likely sites of low-metallicity star formation, hence of binary BH formation.
The analyses of \citet{OSHAUGHNESSY2016} and \citet{Elbert2017} adopt a delay time
distribution function to characterize the merger time of binary BHs and do not determine the properties of
individual binary systems formed in each galaxy. 

In this work, we attempt to overcome the above limitations, and to use individual galaxy properties to predict the number and
properties of binary systems that each galaxy hosts. 

%%%%%%%Fig 1%%%%%
\begin{figure*}
\centering
\includegraphics [width=8.5cm]{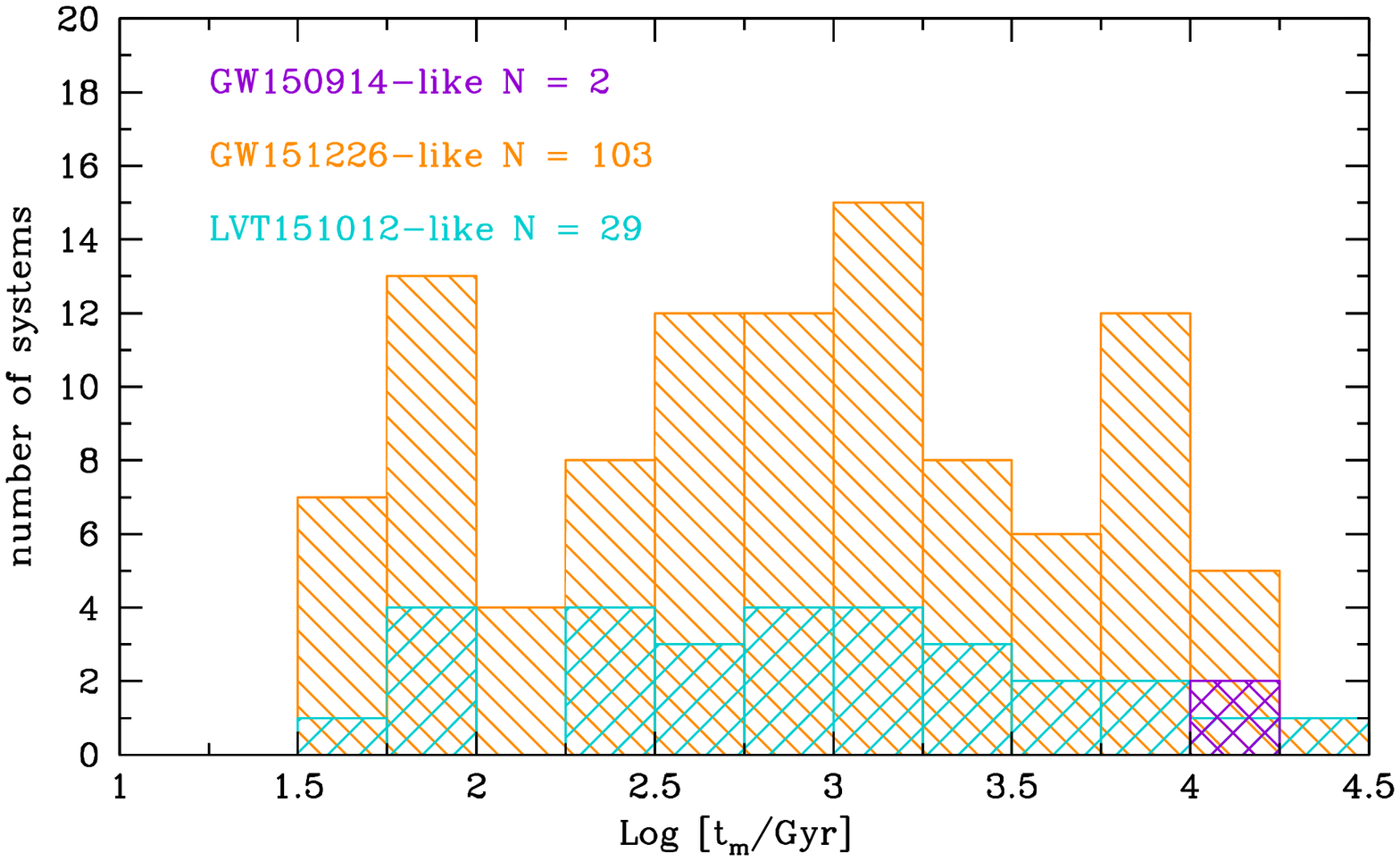}
\includegraphics [width=8.5cm]{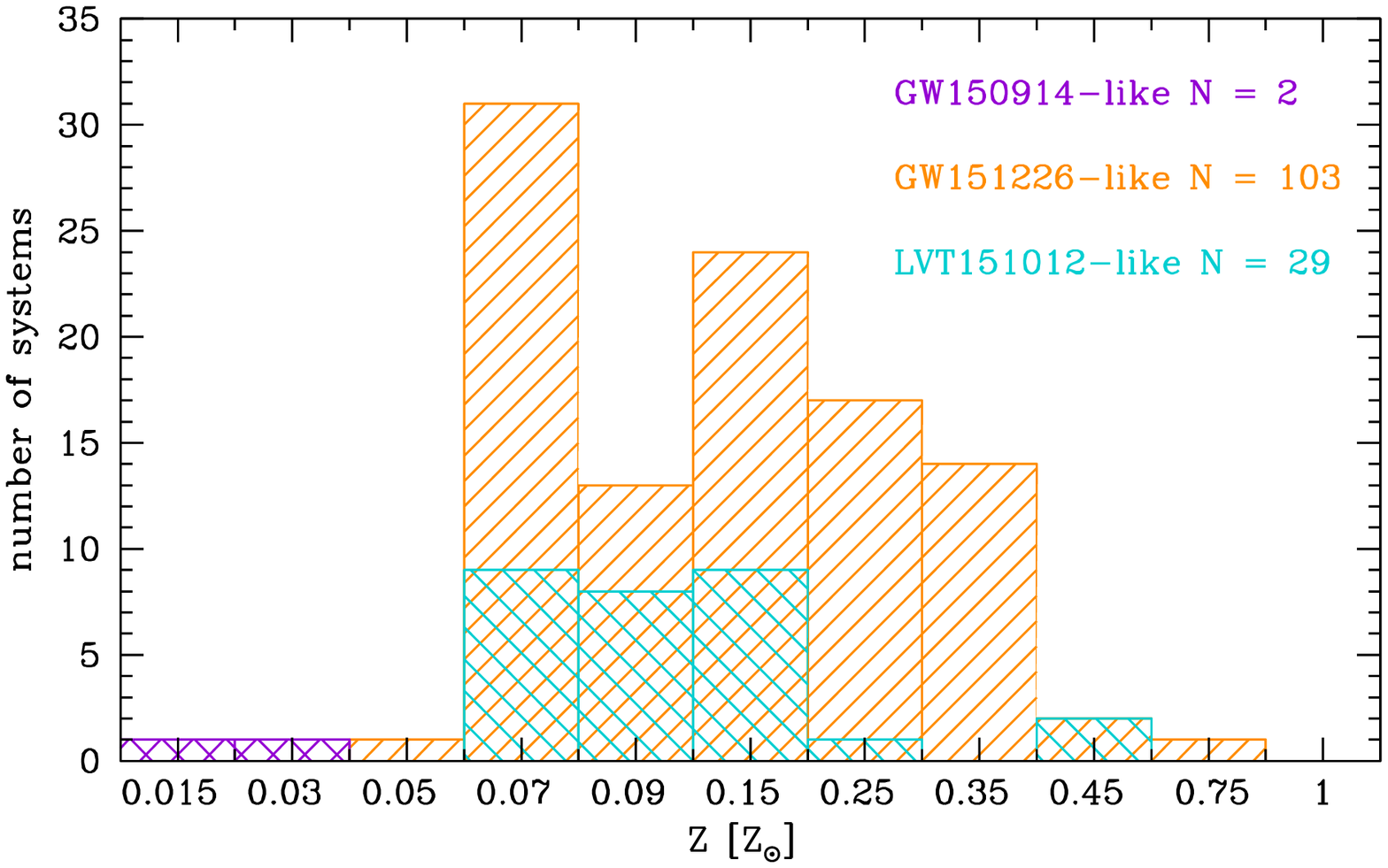}
\caption{The number of events like GW150914, GW151226, and LVT151012 (see the legenda)
predicted by \texttt{SeBa} as a function of metallicity (left panel) and merger time (right panel).}
\label{fig:BHBH}
\end{figure*}
%%%%%%%Fig 1%%%%%

\section{Forming BH-BH systems}
\label{sec:BHBH}

%\subsection{The binary population synthesis model}
\noindent
{\bf The binary population synthesis model:} \\
We adopt the binary population synthesis code \texttt{SeBa}\footnote{http://www.sns.ias.edu/starlab/seba/} 
originally developed by \citet{PZ1996} and \citet{NELEMANS2001}. \texttt{SeBa} follows the evolution of binary systems by taking into 
account the relevant physics involved in their evolution: stellar composition, stellar winds, mass transfer and accretion, magnetic braking, common envelope phase, supernova kicks, and gravitational radiation. 
The original version has been recently modified by \citet{MAPELLI2013}  to include metallicity-dependent prescriptions for stellar evolution, stellar winds, and remnant formation. In particular, metallicity-dependent stellar evolution is based on the polynomial fitting formulas described in \citet{Hurley2000}. Stellar winds were substantially upgraded including 
\citet{VINK2001} prescriptions for O-type massive stars, \citet{Vink2005} formalism for Wolf-Rayet stars, and 
\citet{BELCZYNSKI2010} recipes for luminous blue variable stars. The mass of the compact remnant was derived as described in \citet{MAPELLI2009}: stars with a pre-supernova mass $>40$ M$_\odot$ are assumed to collapse to BH directly, retaining most of their final mass. Binary evolution was modelled as in the default {\tt SeBa} code. For more details on single and binary evolution, see \citet{MAPELLI2013} and  Fig.~1 of \citet{MAPELLI2013} for the derived BH mass spectrum.
We set the binary fraction $f_{\rm bin}=1$ so that all the stars are assumed to be in binary systems. 
Observations show that this assumption is approximately correct for stars with masses $>10 \, M_\odot$  \citep{SANA12}.
For each simulation, we initialize $N= 2 \times 10^{6}$  binaries and follow their evolution with 
initial properties randomly selected from independent distribution functions. 
In particular, the primary stellar mass, $m_{\rm p}$, is distributed according to a Kroupa Initial Mass Function (IMF, \citealt{KROUPA2001}) between $[0.1-100] \, M_{\odot}$. 
The mass of the secondary star, $m_{\rm s}$, is generated according to a flat distribution for the mass ratio $q = m_{\rm s}/m_{\rm p}$
with $0.1 < q \leq 1$. The initial semi-major axis (sma) distribution is flat in log(sma) (see \citealt{PZ1996}) 
ranging from $0.1 \,R_{\odot}$ (Roche lobe contact) up to $10^{6} \, R_{\odot}$. The eccentricity of the binary is
selected from a thermal distribution $f(e) = 2e$ in the $[0 - 1]$ range \citep{HEGGIE1975}.
With this choice of initial conditions, we run simulations for 11 different values of the initial metallicity in the range
$0.01 Z_\odot \leq Z \leq 1 Z_\odot$. 

An extended statistical analysis of the simulations will be presented in a 
forthcoming paper. Here we illustrate the properties of simulated BH-BH systems with primary and secondary masses 
in the range estimated for GW150914 ($m_{\rm BH,p} = 36.2^{+5.2}_{-3.8} \, M_\odot$ and $m_{\rm BH,s} = 29.1^{+3.7}_{-4.4} \, M_\odot$), GW151226 ($m_{\rm BH,p} = 14.2^{+8.3}_{-3.7} \, M_\odot$ and $m_{\rm BH,s} = 7.5^{+2.3}_{-2.3} \, M_\odot$), 
and LVT151012 ($m_{\rm BH,p} = 23^{+18}_{-6} \, M_\odot$ and $m_{\rm BH,s} = 13^{+4}_{-5} \, M_\odot$). In the left panel of Fig.~\ref{fig:BHBH}
we show the number of candidate systems for the three events found in simulated samples with different metallicities.
We find only two massive BH binaries with masses compatible with GW150914 and both require very low initial stellar
metallicities, $Z < 0.05 \, Z_\odot$\footnote{A similar condition ($Z < 0.07 \, Z_\odot$) applies to GW170104-like systems, given the
masses of the primary ($m_{\rm BH,p} = 31.2^{+8.4}_{-6} \, M_\odot$) and secondary  ($m_{\rm BH,s} = 19.4^{+5.3}_{-5.9} \, M_\odot$) black holes.}. GW151226-like systems are more common and their stellar progenitors
have metallicities in the range $0.05 \, Z_\odot \leq Z \leq 0.75 \, Z_\odot$.  Due to the higher mass of the primary BH, 
we find that LVT151012-like systems are less frequent 
and that most of their stellar progenitors have metallicities $0.07 \, Z_\odot \leq Z \leq 0.25 \, Z_\odot$.
In the right panel of Fig.~\ref{fig:BHBH} we show the corresponding distribution of merger times. The two GW150914-like systems are both characterized by long coalescence times $3.87 < {\rm lg}(t_{\rm m}/{\rm Gyr}) <4.12$. 
This is expected as massive BH binaries originate from massive stellar binaries that, given their larger radii, must
have large semi-major axis at birth to avoid merging before the compact binary system forms \citep{Linden2010,
Mapelli2014, Ziosi2014}. Instead, GW151226 and LVT151012-like systems follow
a relatively flat distribution in merger times in the range
$1.37 < {\rm lg}(t_{\rm m}/{\rm Gyr}) \leq 4.25$. Due to the initial mass ratio in the zero age main sequence
($0.4 \leq q < 0.6$), all the GW151226 and LVT151012-like systems 
during their evolution go through a phase of common envelope, which 
drastically reduces the initial orbital separation. For these systems, the flat merger times distribution
 is a consequence of the flat initial distribution of orbital separations. \\

%\subsection{The cosmological simulation}
%\label{subsec:simulation}
\noindent
{\bf The cosmological simulation:}\\ 
We couple the binary information provided by \texttt{SeBa} with the star formation and chemical enrichment histories 
 of all the galaxies contained in a comoving volume $(4 \, \rm cMpc)^3$, experiencing a forming Milky Way (MW)-like galaxy at its center. 
 These
are obtained processing a N-Body cosmological simulation with the \texttt{GAMESH} pipeline \citep{GRAZIANI2015}. The simulation has been
presented in \citet{GRAZIANI2017} and here we briefly summarize the properties relevant to the present analysis.

The N-Body simulation is performed by the code GCD+ \citep{KAWATA2013} and adopts a Planck 2013 cosmology (\citealt{PLANCK2014}, 
$\Omega_{\rm m} = 0.32$,$\Omega_\Lambda=0.68$, $\Omega_{\rm b} = 0.049$ and $h = 0.67$) and initial conditions (ICs) suitable to reproduce a 
MW-like halo in the present Universe, starting from a cosmological volume of 
about (83.53 cMpc)$^3$. 
Once a MW-sized halo is identified at the largest scale, we re-create ICs for a multi-resolution, zoom-in simulation resolving the same halo 
with mass $M_{\rm MW} = 1.7 \times 10^{12} M_\odot$ at the center of a volume of (4 cMpc)$^3$ with a dark matter particle resolution mass of 
$3.4 \times 10^5 M_\odot$. Hereafter we will refer to this volume as the Local Group (LG) of the MW-like halo.
The simulation runs from $z = 20$ to $z =0$ and we store the outputs with a time resolution of $15$~Myr at $z > 10$, and of $100$~Myr at $z \leq 10$. 
In each snapshot, halos are identified by a standard FoF algorithm adopting linking parameter of $b = 0.2$ and a minimum number of particles of 100.
A particle-based merger tree (MT) has been computed to exactly establish each halo ancestor/descendant relationships, also accounting for all the 
dynamical processes regulating halo assembly: accretion, mergers, tidal stripping and halo disruption. The simulation is in good agreement with observations 
and with independent theoretical studies \citep{GRAZIANI2017}. At all redshifts, the resulting halo mass function in the LG
matches the prediction of the analytic Press-Schecter distribution for all halos with $M \leq 10^{10} M_\odot$. Below $z \sim 3$, the 
central halo dynamically dominates the LG region (half of the MW halo mass is already assembled by $z \sim 1.5$) and 
halos with mass $M > [10^{10} - 10^{11}] ~M_\odot$ are found to evolve faster than expected for an average cosmic region.

The baryonic evolution of galaxies is performed by \texttt{GAMESH} accounting for star formation, metal enrichment, Pop\ III/Pop\ II transition and supernova (SN)-driven feedback. Radiative feedback is modeled by adopting an instant reionization prescription which suppresses star formation in all galaxies hosted by mini-halos (with virial temperatures $T_{\rm vir} < 10^4$~K) found at $z \leq 6$. By calibrating the model free parameters to reproduce the stellar, gas and metal mass observed in the MW, we find the simulated galaxies to be in good agreement with recent observations of candidate MW progenitors at $0 < z < 2.5$, with the galaxy main sequence, mass-metallicity relation and fundamental plane of metallicity relations in the redshift range $0 < z < 4$. 

It is worth pointing out that although the simulated cosmic volume is very small compared to the current instrumental range of the advanced LIGO detector ($d \leq 1$~Gpc), the adopted resolution allows us to simulate DM structures down to a minimum mass  of $\sim 3.4 \times 10^7 M_\odot$, where star formation in low-metallicity environments (hence massive BH formation) is more likely to occur. \\

\begin{figure*}
\centering
\includegraphics [width=8.5cm]{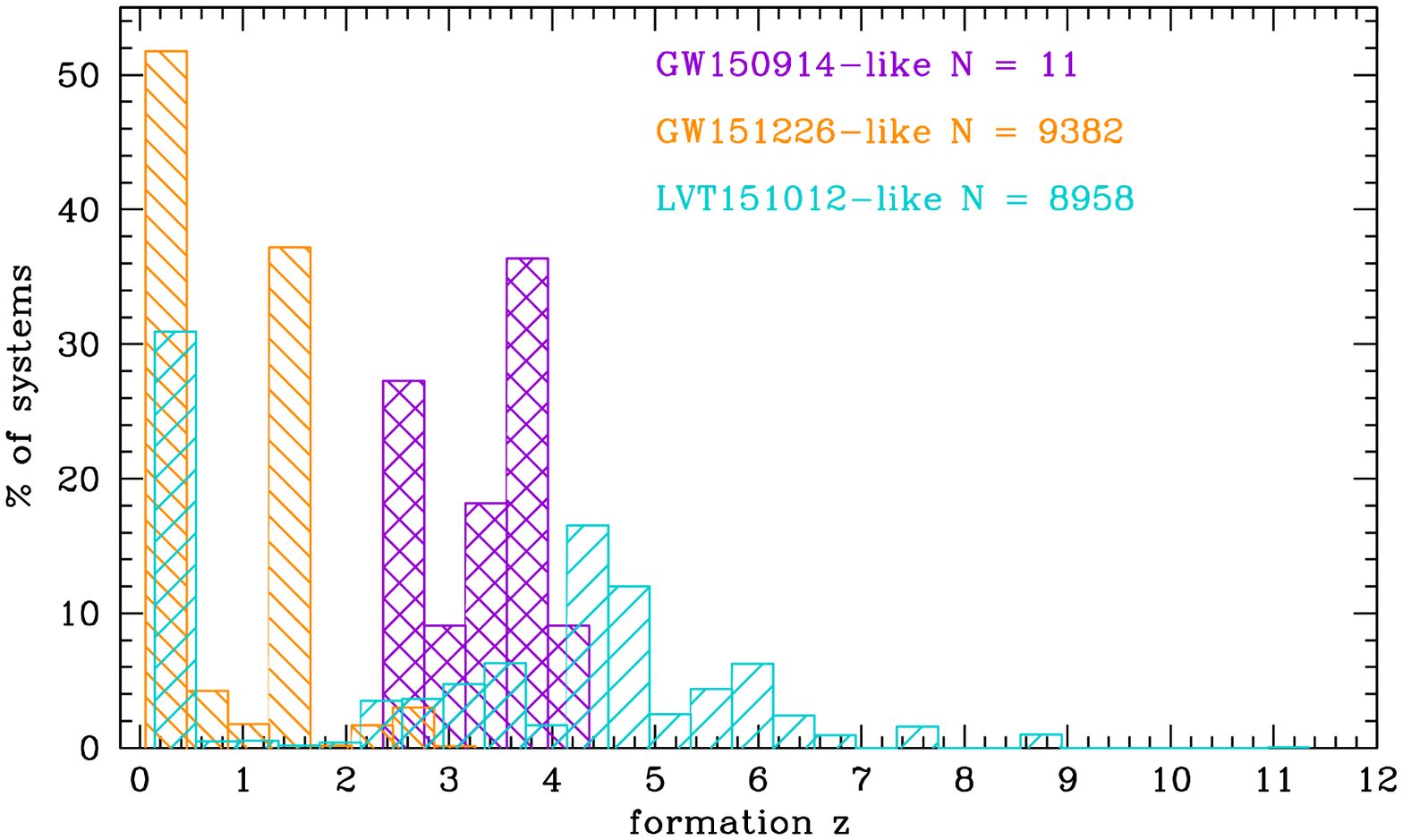}
\includegraphics [width=8.5cm]{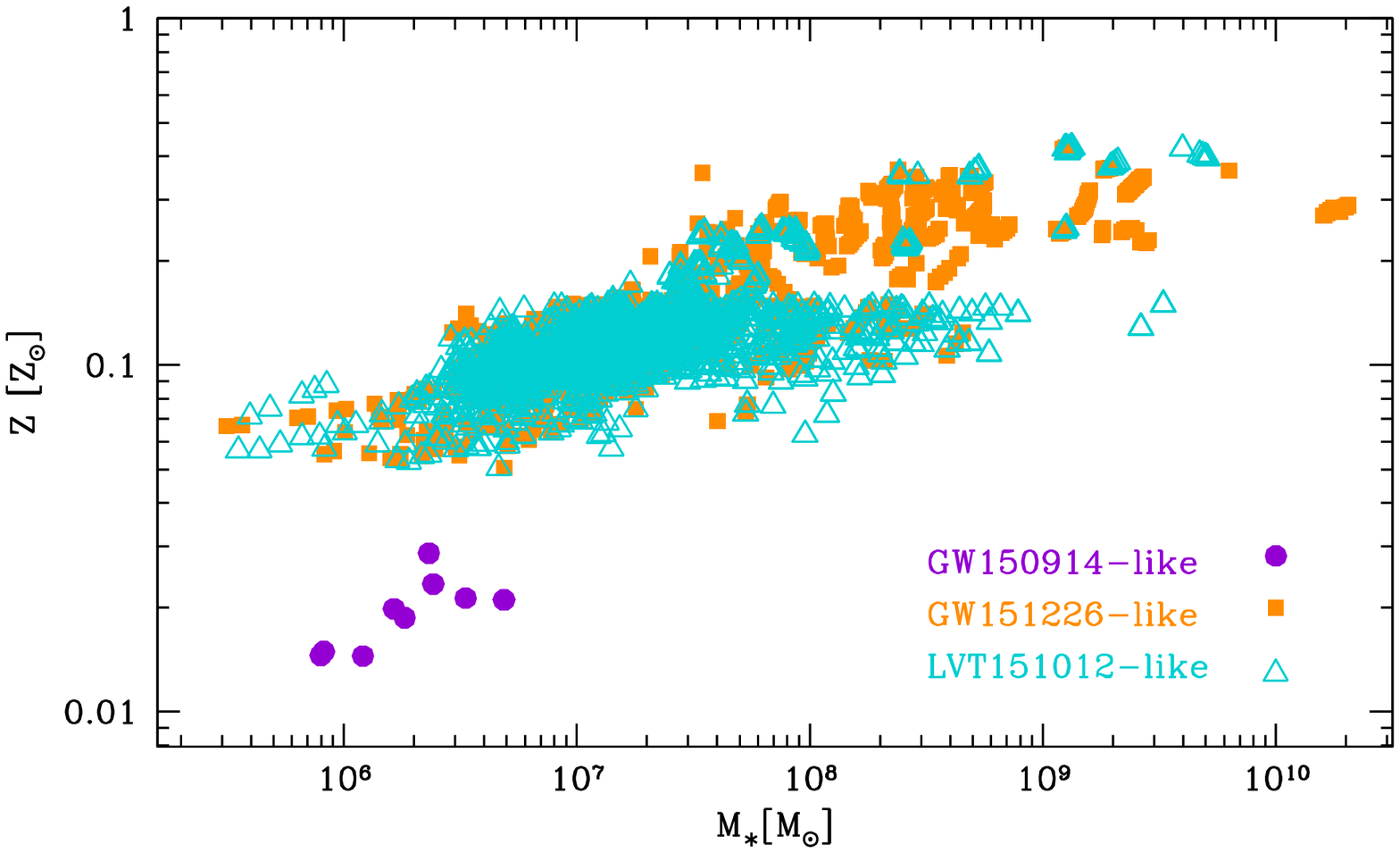}
\caption{Properties of the formation sites of GW150914, GW151226 and LVT151012-like systems in the simulation. {\bf Left panel}: percentage
of systems as a function of formation redshift. {\bf Right panel:} gas metallicity as a function of the stellar mass of the galaxies where
the simulated LIGO-like events are formed. }
\label{fig:formation}
\end{figure*}

\begin{figure}
\centering
\includegraphics [width=\hsize]{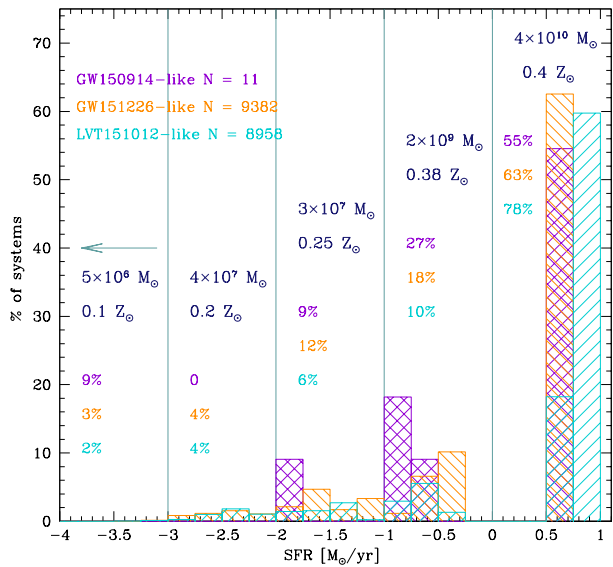}
\caption{Properties of the coalescence sites of GW150914, GW151226 and LVT151012-like systems in the simulation. Histograms
show the percentage of systems hosted in galaxies binned by their star formation rate at the time of coalescence. The 
percentage of BH binaries hosted in galaxies with SFRs within the broader bins identified by the vertical lines are also shown,
together with the average host  stellar mass and metallicity. The numbers in the leftmost bin refer to BH binaries hosted 
in galaxies with negligible or null star formation activity, as indicated by the arrow (see text).}
\label{fig:merger}
\end{figure}
\noindent
{\bf BH binary formation in the simulated galaxies:} \\
In order to determine the properties of compact binaries formed in each galaxy,
we assume that the stellar progenitors have the metallicity of the gas in which they form and we randomly extract from 
the \texttt{SeBa} output with the closest metallicity a number of binary systems until we reach the total mass of newly formed stars predicted by the cosmological simulation.
As a result of this procedure, we can predict the metallicity-dependent formation and merger rates of compact binaries 
and the properties of compact binary systems hosted in each galaxy of the  LG at any given time from $0 \leq z \leq 20$.

\section{Results}
\label{sec:results}
We select, among all the simulated binaries,
systems with properties consistent with GW150914, GW151226 and LVT151012. This is done by requiring that the system has primary and secondary
BH masses in the observed range and that it coalesces at $z_{\rm m} = 0.09^{+0.03}_{-0.04}$ for GW150914 and GW151226, 
and at $z_{\rm m} = 0.20^{+0.09}_{-0.09}$ for LVT151012. We find  $N = 11, 9382$, and $8958$ 
for GW150914, GW151226, and LVT151012, 
corresponding to merger rates, in the LG, of $\sim 0.2, 162$, and $70$ Gpc$^{-3}$yr$^{-1}$, respectively.
%$\sim 0.3, 200$, and $84$ Gpc$^{-3}$yr$^{-1}$, respectively. 
These differ from the rates obtained by 
\citet{ABBOTT2016_BHBHO1} based on populations with masses matching the observed events and assuming a uniform distribution
in comoving volume\footnote{\citet{ABBOTT2016_BHBHO1} find $3.4^{+8.8}_{-2.8}, 36^{+95}_{-30}$, and 
$9.1^{+31}_{-8.5}$ Gpc$^{-3}$yr$^{-1}$ for GW150914, GW151226, and LVT151012, 
where we quote the median values with 90\% confidence from their Table II.}. Indeed, for this comparison to
be meaningful a larger simulation box probing a more representative cosmological volume should be considered
(see \citealt{MAPELLI2017} for a similar approach using the Illustris simulation). 

For each class of systems, we reconstruct from the simulation output the properties of their
formation and coalescence sites. Fig.~\ref{fig:formation} shows the redshift distribution (left panel) and the mass-metallicity relation (right panel) of the 
host galaxies at BH binary formation. The redshift range over which the systems formed reflects the
corresponding merger time distribution. GW150914-like systems are characterized by long merger times and form in the range
$2.36 \leq z_{\rm f} \leq 4.15$. GW151226 and LVT151012-like systems, which have a broad range of possible merger times,
can form in relatively wide redshift intervals $0.06 \leq z_{\rm f} \leq 2.97$ and $0.14 \leq z_{\rm f} \leq 11.3$, respectively. However,
we find that while only 6\% of GW151226-like systems have $z_{\rm f} > 2$, among LVT151012 systems, 67\% have $z_{\rm f} > 2$, 48\%
have $z_{\rm f} > 4$, and 6\% formed in the pre-reionization epoch, with $z_{\rm f} > 6$. 
In addition, all the GW150914-like systems are born 
in low-metallicity dwarfs with stellar
mass $M_{\ast} < 5 \times 10^6 M_\odot$, while 88\% (70\%) of GW151226 (LVT151012)-like systems form in galaxies with $M_\ast > 10^8 M_\odot$.
%These percentages do not necessarily reflect the density of points in the right panel of Fig.~\ref{fig:formation} as the latter represent host
%galaxies where multiple BH binary systems may form, especially for larger stellar masses.

We track each GW150914, GW151226 and LVT151012-like BH system from formation to coalescence using the particle-based MT
and we identify the galaxy where it resides when the merger event occurs. The results of this procedure are illustrated in Fig.~\ref{fig:merger}.
The histograms represent the percentage of systems hosted in galaxies characterized by their star formation rate (SFR) at the time of coalescence.
The numbers shown in the plot refer to the percentage of systems in larger SFR intervals as identified by the vertical lines. For each of these
intervals, we also report the average stellar mass and metallicity of the hosts. The numbers  in the leftmost bin refer to systems 
which are not forming stars at the time of coalescence.
Our results suggest
that GW150914, GW151226 and LVT151012-like systems have the largest probability to be hosted in galaxies with masses
$M_{\ast} \sim 4 \times 10^{10} M_\odot$, metallicities $Z \sim  0.4 \, Z_{\odot}$ and which form stars at moderate rates $\sim 5 M_\odot$/yr. 
These are the progenitors of the MW-like galaxy at the redshift of coalescence \citep{GRAZIANI2017}.
Smaller galaxies, with stellar masses encompassing the range of LMC, SMC and dwarfs, have a smaller probability to have 
hosted the first three GW events. 

Finally, we find a small but not negligible percentage of GW150914-like systems ($\sim$ 10\%) whose coalescence could be hosted
in very small galaxies, with $M_\ast \sim 5 \times 10^6 M_\odot$, where star formation has been suppressed by radiative feedback.  These BH
binaries have formed in dwarf galaxies at $z \sim 4$ which, as a consequence of  stripping events, lower their (dark and baryonic) mass
below the limit that renders gas infall unfavorable (virial temperatures smaller than $\sim 2 \times 10^4$~K) in the rising UV background accompanying
reionization. Although these findings are based on a very small number statistics (1 out of 11 GW150914-like systems),  
they suggest that the hosts of massive BH binary mergers might be similar to the least massive among dwarf spheroidals or to the more massive among ultra-faint galaxies. 

\section{Conclusions}
\label{sec:conclusions}

The work that we have presented is only a first step towards a more extended analysis of different classes of compact
binary systems and their detectability as gravitational wave sources. 
Ultimately, our goal is to predict the most probable sites of formation and coalescence of different classes of compact
binary systems. This information will help  the search for the host galaxy properties and to identify the
electromagnetic counterparts of some gravitational wave sources. 

While we plan to present an extended analysis in future work, here we select, among all the simulated binaries, GW150914, GW151226 and LVT151012-like systems. 
We find that, despite the fact that these systems require
different galaxy properties at the time of their formation, they all have a larger probability to be hosted
in massive, star forming galaxies at the time of their coalescence. This is in agreement with what was recently suggested by
independent studies \citep{LAMBERTS2016, Elbert2017}. However, massive BH binaries which preferentially form in metal-poor dwarfs at $2.5 \leq z \le 4.2$,
can also fail to be incorporated in more massive galaxies during hierarchical evolution. As a result, they might have evolved within galaxies which are fragile to radiative 
feedback effects, reaching coalescence in small non star forming dwarf satellites.

\section*{Acknowledgments}
The research leading to these results has received funding from the European Research Council under the Grant Agreement n. 306476. MM and MS acknowledge financial support from the Italian Ministry of Education, University and Research (MIUR) through grant FIRB 2012 RBFR12PM1F. MM acknowledges financial support from the MERAC Foundation.

\bibliographystyle{mn2e}
\bibliography{GWbinaries}

\label{lastpage}

\end{document}